\documentclass[conference]{IEEEtran}
\IEEEoverridecommandlockouts
\usepackage{cite}
\usepackage{amsmath,amssymb,amsfonts,bm,mathtools,makecell,cite,stfloats}
\usepackage{algorithmic}
\usepackage{graphicx}
\usepackage{textcomp}
\usepackage{xcolor}
\def\BibTeX{{\rm B\kern-.05em{\sc i\kern-.025em b}\kern-.08em
    T\kern-.1667em\lower.7ex\hbox{E}\kern-.125emX}}
\begin{document}

\title{Distributed Weighted Least Squares Estimator Based on ADMM\\
}


\author{
	\IEEEauthorblockN{Shun~Liu\textsuperscript{1,2},  
		Zhifei~Li\textsuperscript{3}, Weifang~Zhang\textsuperscript{4}
		Yan~Liang\textsuperscript{1,2}} 	
	\IEEEauthorblockA{\textsuperscript{1} School of Automation, Northwestern Polytechnical University, Xi'an, China} 
	\IEEEauthorblockA{\textsuperscript{2} Key Laboratory of Information Fusion Technology, Ministry of Education, Xi'an, China} 
	\IEEEauthorblockA{\textsuperscript{3} School of Space Information, Space Engineering University, Beijing, China}
	\IEEEauthorblockA{\textsuperscript{4} Shanghai Electro-Mechanical Engineering Institute, Shanghai, China}
	\IEEEauthorblockA{Email: liushun\_ls@mail.nwpu.edu.cn, lizhifei17@nudt.edu.cn, fantasticfang@sina.com, liangyan@nwpu.edu.cn} 
}

\maketitle

\begin{abstract}
Wireless sensor network has recently received much attention due to its broad applicability and ease-of-installation. This paper is concerned with a distributed state estimation problem, where all sensor nodes are required to achieve a consensus estimation. The weighted least squares (WLS) estimator is an appealing way to handle this problem since it does not need any prior distribution information. To this end, we first exploit the equivalent relation between the information filter and WLS estimator. Then, we establish an optimization problem under the relation coupled with a consensus constraint. Finally, the consensus-based distributed WLS problem is tackled by the alternating direction method of multiplier (ADMM). Numerical simulation together with theoretical analysis testify the convergence and consensus estimations between nodes.
\end{abstract}


\section{Introduction}
Recently, distributed estimation schemes have gained immense popularity due to their flexibility for large networks and high fault tolerance \cite{b1,b2}. Compared with the centralized fashion, it does not need a fusion center. So it is capable of overcoming congestion of massive data. In view of this, distributed estimation has been used in a wide range of domains, e.g., environmental monitoring, surveillance, cooperative target tracking \cite{b3}.  

The objective of a distributed estimation fashion is to achieve a consensus estimation between nodes using only local noisy measurements and information from one-hop neighbors. To this end, the existing algorithms rely on different consensus strategies, e.g., average consensus (AC) scheme \cite{b4}, diffusion strategy \cite{b5}, minimizing weighted Kullback-Leibler divergence \cite{b6}, and covariance intersection fusion rule \cite{b7}. Moreover, Wang et al. constructed an equivalent consensus problem from maximum-a-posterior (MAP) perspective \cite{b8}. And then the problem with a consensus constraint is handled via the alternating direction method of multiplier (ADMM) technique. A main restriction in \cite{b8} is it require the posterior distribution function w.r.t estimated quantity. However, this requirement cannot always be satisfied as it needs exact prior knowledge to hold a conjugate prior structure. 

Different from the aforementioned consensus schemes, this work considers the consensus problem via weighted least squares (WLS) viewpoint \cite{b9}. Compared with MAP estimator, a unique requirement for WLS criterion is the first and second moment information about noise (error). Obviously, this merit is appealing because the practical prior knowledge is limited instead of sufficient. Once the prior is not available, the related filters based on MAP will be invalid. To derive a distributed filter within WLS criterion, we first establish an equivalent relation between the information filter and WLS estimator. Next, to guarantee a consensus-based solution between nodes, we construct a novel optimization function and add consensus constraints to force these variables on neighboring nodes to reach an agreement. Finally, the optimization problem is addressed by ADMM technique. Numerical simulations together with theoretical analyses testify the effectiveness of the proposed filter on node consensus and convergence.

\section{Problem Formulation} \label{sec:2}
We consider a discrete-time linear dynamic system with
the state vector $\bm x _k \in \mathbb R^m$. At each scan time $k$, $\bm x_k$ is observed over the sensor network $\mathcal{G} = \{ \mathcal{J}, \mathcal{A} \}$, where $\mathcal{J}:=\{1,\cdots,J\} $ is the set of sensor nodes, $\mathcal{A \subseteq J \times J}$ is the set of edges such that $(s, j) \in \mathcal{A}$ if node $s$ can communicate with $j$. For each node $s \in \mathcal{J}$, $\mathcal J^s$ denotes the set of its neighboring nodes (excluding $s$ itself), i.e., $\mathcal J^s := \{j: (j,s) \in \mathcal{A}, s \neq j\}$, and let $J_s$ be the number of the
neighboring nodes of node $s$. The accumulated data from all sensor nodes at time $k$ is denoted as $\mathcal Y_k = \{\bm y_{k,s}\}_{s \in \mathcal{S}}$. The dynamic
process and the local measurement model on each node $s \in \mathcal{J}$ is given as follows:
\begin{equation} \label{eq1}
\bm x_{k+1} = \mathbf F_k \bm x_k + \bm w_k,
\end{equation}
\begin{equation} \label{eq2}
\bm y_{k,s} = \mathbf H_{k,s} \bm x_k + \bm v_{k,s},
\end{equation}
where $\mathbf F_k \in \mathbb R^{m \times m}$ and $\mathbf H_{k,s} \in \mathbb R^{n \times m}$ are the state transition and measurement matrices, respectively, $\bm w_k \in \mathbb R^m$ and $\bm v_{k,s} \in \mathbb R^n$ are the process and measurement noise sequences, respectively. 
Here, we assume that zero-mean noises $\bm w_k$ and $\bm v_{k,s}$ with covariance matrices $\mathbf Q_k$ and $\mathbf R_{k,s}$ are mutually uncorrelated, i.e.,
\begin{equation} \label{eq3}
\mathbb{E} \left\{  \begin{bmatrix}
\bm w_k \\ \bm v_{k,s}
\end{bmatrix} \begin{bmatrix}
\bm w_p \\ \bm v_{p,j}
\end{bmatrix}^{\mathsf{T}} \right\} =
\begin{bmatrix}
\mathbf Q_k & \bm 0 \\ \bm 0 & \mathbf R_{k,s}
\end{bmatrix} \delta_{s,j} \delta_{k,p},
\end{equation} 
where $\delta_{s,j}$ is the Kronecker delta function, i.e., $\delta_{s,j} = 1$ if $s=j$. 

We further define the collective measurement, measurement noise, measurement matrix, and measurement noise covariance as $\bm y_k:=[\bm y_{k,1}^{\mathsf{T}}, \cdots, \bm y_{k,J}^{\mathsf{T}}]^{\mathsf{T}}$, $\bm v_k:=[\bm v_{k,1}^{\mathsf{T}}, \cdots, \bm v_{k,J}^{\mathsf{T}}]^{\mathsf{T}}$, $\mathbf H_k:=[\mathbf H_{k,1}^{\mathsf{T}}, \cdots, \mathbf H_{k,J}^{\mathsf{T}}]^{\mathsf{T}}$, and $\mathbf R_k := \mathrm{blkdiag} [\mathbf R_{k,1},\cdots,\mathbf R_{k,J}]$, respectively. Then, \eqref{eq2} is rewritten as an integrated form 
\begin{equation} \label{eq4}
\bm y_k = \mathbf H_k \bm x_k + \bm v_k, 
\end{equation}
where $\mathbb{E} (\bm v_k) = [\bm 0^{\mathsf{T}}, \cdots, \bm 0^{\mathsf{T}}]^{\mathsf{T}}$, $\mathrm{Cov} (\bm v_k) =  \mathbf R_k$.

Since the probability distribution knowledge about noise is unknown, we cannot anchor our hope on MAP criterion. Thus, the objective of the work is to make the estimation $\hat{\bm x}_{k,s}$ on each node $s \in \mathcal{J}$ achieve a consensus by resorting WLS criterion. Meanwhile, the estimation $\hat{\bm x}_{k,s}$ should approach to the centralized estimation $\hat{\bm x}_{k}$ as much as possible. 
\section{Distributed WLS Estimator}  
This section first provides a relation between the information filter and WLS estimator. Then, under the relation, an optimization problem with consensus constraints is constructed. 
\subsection{Information Filter and Its Relation to WLS Estimator}
Under the collective system models as shown in \eqref{eq1} and \eqref{eq4}, centralized information filter (CIF) gives an optimal estimation within the linear minimum mean-square-error (LMMSE) criterion \cite{b10}. The CIF consists of two steps, namely, measurement update and time update procedure. Define the estimation errors $\tilde{\bm x}_{k|k-1}:=\hat{\bm x}_{k|k-1} - {\bm x}_k$ and $\tilde{\bm x}_{k|k}:=\hat{\bm x}_{k|k} - {\bm x}_k$ and the corresponding error covariances as $\mathbf P_{k|k-1} := \mathbb{E} \left( \tilde{\bm x}_{k|k-1} (\tilde{\bm x}_{k|k-1})^{\mathsf{T}}\right) $ and $\mathbf P_{k|k} := \mathbb{E} \left( \tilde{\bm x}_{k|k} (\tilde{\bm x}_{k|k})^{\mathsf{T}}\right) $, respectively. Let information matrix of the error covariance $\mathbf P_{k|k}$ as $\mathbf{\Omega}_{k|k} := \left(\mathbf P_{k|k}\right)^{-1}$. When the initial estimation $\hat{\bm x}_{0|0}$ and information matrix $\mathbf{\Omega}_{0|0}$ are given, CIF can obtain a recursive solution over time evolution as follows:
\begin{itemize}
	\item Measurement Update
	\begin{equation} \label{eq5}
	 \mathbf{\Omega}_{k|k} = \mathbf{\Omega}_{k|k-1} + \mathbf H_k^{\mathsf{T}} \mathbf R_k^{-1} \mathbf H_k,
	\end{equation}
	\begin{equation} \label{eq6}
	\begin{split}
	\hat{\bm x}_{k|k} = & \left( \mathbf{\Omega}_{k|k-1} + \mathbf H_k^{\mathsf{T}} \mathbf R_k^{-1} \mathbf H_k \right)^{-1} \\
	& \times \left( \mathbf{\Omega}_{k|k-1} \hat{\bm x}_{k|k-1} + \mathbf H_k^{\mathsf{T}} \mathbf R_k^{-1} \bm y_k \right),
	\end{split}
	\end{equation} 
	\item Time Update
	\begin{equation} \label{eq7}
	\hat{\bm x}_{k|k-1} = \mathbf F_k \hat{\bm x}_{k-1|k-1},
	\end{equation}
	\begin{equation} \label{eq8}
	\mathbf{\Omega}_{k|k-1} = \left( \mathbf F_k \left( \mathbf{\Omega}_{k-1|k-1} \right)^{-1} (\mathbf F_k)^{\mathsf{T}} + \mathbf Q_k \right)^{-1}.
	\end{equation}
\end{itemize}  

Combined \eqref{eq4} with error $\tilde{\bm x}_{k|k-1}$, the measurement update procedure in CIF is expressed as an equivalent form:
 \begin{equation} \label{eq9}
 \begin{bmatrix}
 \hat{\bm x}_{k|k-1} \\ \bm y_{k}
 \end{bmatrix} =
 \begin{bmatrix}
 \mathbf I \\ \mathbf H_{k}
 \end{bmatrix} \bm x_k + 
 \begin{bmatrix}
 \tilde{\bm x}_{k|k-1} \\ \bm {v}_{k}
 \end{bmatrix}.
 \end{equation}
Letting 
\begin{equation*}
\bm {\mathcal Y}_k =  \begin{bmatrix}
\hat{\bm x}_{k|k-1} \\ \bm y_{k}
\end{bmatrix} ,\: \bm {\mathcal H}_k = \begin{bmatrix}
\mathbf I \\ \mathbf H_{k}
\end{bmatrix} ,\:
\bm {\beta}_k =  \begin{bmatrix}
\tilde{\bm x}_{k|k-1} \\ \bm {v}_{k}
\end{bmatrix},
\end{equation*}
we have 
 \begin{equation} \label{eq10}
\bm {\mathcal Y}_k =  \bm {\mathcal H}_k \bm x_k + \bm {\beta}_k,
 \end{equation} 
where $\mathbb{E} (\bm {\beta}_k) = [\bm 0^{\mathsf{T}}, \bm 0^{\mathsf{T}}]^{\mathsf{T}}$, $\mathrm{Cov} (\bm {\beta}_k) =  \bm {\mathcal C}_k$.
 
Equation \eqref{eq10} is a standard WLS form, where the optimal wight matrix is $\bm {\mathcal C}_k^{-1} = [\mathbf{\Omega}_{k|k-1},\mathbf R_{k}^{-1}]$. According to the WLS criterion, a WLS solution of \eqref{eq10} is
\begin{align} 
 &\hat{\bm x}_{k|k}^{\mathrm{WLS}} =\arg \underset{\bm x_k}{\min} \left( \bm {\mathcal Y}_k - \bm {\mathcal H}_k \bm x_k \right)^{\mathsf{T}} \bm {\mathcal C}_k^{-1}
\left( \bm {\mathcal Y}_k - \bm {\mathcal H}_k \bm x_k \right) \notag \\
& = \arg \underset{\bm x_k}{\min} \left(   \Vert \bm y_k - \mathbf H_k \bm x_k \Vert^2_{\mathbf R_{k}^{-1}} + \Vert \bm x_k - \hat{\bm x}_{k|k-1} \Vert^2_{\mathbf{\Omega}_{k|k-1}}  \right), \label{eq12}
\end{align}
where $\Vert \bm a  \Vert^2_{\mathbf G} = \bm a^{\mathsf{T}} \mathbf G \bm a $ denotes Mahalanobis norm.

Since the cost function \eqref{eq12} is convex on $\bm x_k$, taking the derivative of $\bm x_k$ on the right side of \eqref{eq12} yields a WLS estimation and its corresponding information covariance 
\begin{subequations} \label{eq13}
	\begin{equation} \label{eq13a}
	\begin{split}
	\hat{\bm x}_{k|k}^{\mathrm{WLS}} =  & \left( \mathbf{\Omega}_{k|k-1} + \mathbf H_k^{\mathsf{T}} \mathbf R_k^{-1} \mathbf H_k \right)^{-1} \\
	& \times \left( \mathbf{\Omega}_{k|k-1} \hat{\bm x}_{k|k-1} + \mathbf H_k^{\mathsf{T}} \mathbf R_k^{-1} \bm y_k \right), 
	\end{split}
	\end{equation}
	\begin{equation} \label{eq13b}
	\mathbf{\Omega}_{k|k}^{\mathrm{WLS}} = \mathbf{\Omega}_{k|k-1} + \mathbf H_k^{\mathsf{T}} \mathbf R_k^{-1} \mathbf H_k.
	\end{equation}
\end{subequations} 
The results in \eqref{eq13} equal to that shown in \eqref{eq5} and \eqref{eq6}. In other words, there exists an equivalent relation between the WLS estimator and centralized information filter.

\subsection{Distributed WLS Estimator}
In this section, we aim to achieve \eqref{eq13} in a distributed manner where every node only exchanges information with its one-hop neighboring nodes. The main difficulty in doing this is that how to accomplish a consensus between nodes. To solve the difficulty, we decompose \eqref{eq12} into $J$ local parallel optimization problems and add consensus constraints $\hat{\bm x}_{k,s} = \hat{\bm x}_{k,j}, \, \forall s \in \mathcal J,\, j \in \mathcal J^s$ \cite{b8}. That is, the problem is transformed into the following form
\begin{align}
&{}\arg \underset{\{  {\bm x}_{k,s} \}}{\min} \sum_{s=1}^{J} f_s(\bm x_{k,s}), \notag  \\
&{} \mathrm{s.t.}\quad {\bm x}_{k,s} = {\bm x}_{k,j}, \, \forall s \in \mathcal J,\, j \in \mathcal J^s, \label{eq14}
\end{align}
with 
\begin{equation} \label{eq15}
\begin{split}
f_s(\bm x_{k,s}) = & \Vert \bm y_{k,s} - \mathbf H_{k,s} \bm x_{k,s} \Vert^2_{\mathbf R_{{k,s}}^{-1}} \\
& + \frac{1}{J}  \Vert \bm x_{k,s} - \hat{\bm x}_{k|k-1,s} \Vert^2_{\mathbf{\Omega}_{k|k-1,s}}.
\end{split}
\end{equation}
We employ the ADMM technique \cite{b11,b12} to address the constrained optimization problem shown in \eqref{eq14}. To update the variables $\bm x_{k,s}$ in parallel among all nodes, a auxiliary variable $\bm z_{s,j} $ is introduced, and then the problem in  \eqref{eq14} is rewritten as 
\begin{align}
&{}\arg \underset{\{ {\bm x}_{k,s} \} \{ {\bm z}_{s,j} \} }{\min} \sum_{s=1}^{J} f_s(\bm x_{k,s}), \notag  \\
&{} \mathrm{s.t.}\quad {\bm x}_{k,s} = {\bm z}_{s,j}, {\bm z}_{s,j} = {\bm x}_{k,j}, \, \forall s \in \mathcal J,\, j \in \mathcal J^s. \label{eq16}
\end{align}
In \eqref{eq16}, the auxiliary variable $\bm z_{s,j} $ decouples local variable ${\bm x}_{k,s}$ and those variables $\{\bm x_{k,j} \}_{j \in \mathcal J^s}$ from its neighboring nodes $j \in \mathcal J^s$. Let $\bm \lambda_{sj1} (\bm \lambda_{sj2})$ denotes the Lagrange multiplier corresponding to the constraint ${\bm x}_{k,s} = {\bm z}_{s,j} ({\bm z}_{s,j} = {\bm x}_{k,j})$. Then, the augmented Lagrangian function for the problem \eqref{eq16} is given as follows
\begin{align} \label{eq17}
& \mathcal{L}_{\rho}\left(\left\{{\bm x}_{k,s}\right\},\left\{\bm z_{s,j}\right\},\{\bm \lambda_{sj1},\bm \lambda_{sj2} \}\right)= \notag \\
& \frac{1}{2} \sum_{s=1}^{J}\left(f_s(\bm x_{k,s})+\rho \sum_{j \in \mathcal{J}^{s}}\left\|{\bm x}_{k,s} - {\bm z}_{s,j}+\frac{\bm \lambda_{sj1}}{\rho}\right\|_{F}^{2}\right. \notag \\
& \left.+\rho \sum_{j \in \mathcal{J}^{s}}\left\|{\bm z}_{s,j} - {\bm x}_{k,j} + \frac{\bm \lambda_{sj2}}{\rho}   \right\|_{F}^{2}\right),
\end{align}
where $\rho >0$ is a penalty parameter \cite{b13}.

We employ the ADMM technique to tackle the problem \eqref{eq17} in a cyclic way by minimizing $\mathcal{L}_{\rho}$ w.r.t local variables $\{ {\bm x}_{k,s} \}$ and auxiliary variables $\{ \bm z_{s,j} \}$. As for the dual variables $\{\bm \lambda_{sj1},\bm \lambda_{sj2} \}$, one can use the gradient ascent step to update their values. Taking
the derivative of the Lagrangian function \eqref{eq17} w.r.t each variable,
and setting the corresponding derivative to zero, yields 

\newcounter{MYtempeqncnt}
\begin{figure*}[b]
	\normalsize
	\hrulefill	
	\setcounter{MYtempeqncnt}{\value{equation}}
	\setcounter{equation}{16}
	\begin{align} \label{eq18}
	\bm x_{k,s}^{l} = 	& \left[ 2 \rho J_s \mathbf I + \mathbf H_{k,s}^{\mathsf{T}} \mathbf R_{k,s}^{-1} \mathbf H_{k,s} + \frac{1}{J} \mathbf{\Omega}_{k|k-1,s}\right] ^{-1} 
	 \notag \\
	& \times \left[  \mathbf H_{k,s}^{\mathsf{T}} \mathbf R_{k,s}^{-1} \bm y_{k,s} + \frac{1}{J} \mathbf{\Omega}_{k|k-1,s} \hat{\bm x}_{k|k-1,s} 
	+ \sum_{j \in \mathcal{J}^{s}} \left(  \bm \lambda_{js2}^{l-1} - \bm \lambda_{sj1}^{l-1} +  \rho (\bm z_{s,j}^{l-1} + \bm z_{j,s}^{l-1}) \right)  \right],  
	\end{align}
	\setcounter{equation}{\value{MYtempeqncnt}}
	\vspace*{4pt}
\end{figure*}
\setcounter{equation}{17}

\begin{equation} \label{eq19}
\bm z_{s,j}^{l} = \frac{1}{2 \rho} \left(  \bm \lambda_{sj2}^{l-1} - \bm \lambda_{sj1}^{l-1}  \right) + \frac{1}{2} \left( \bm x_{k,s}^{l} + \bm x_{k,j}^{l}  \right), 
\end{equation}
\begin{equation} \label{eq20}
\bm \lambda_{sj1}^{l} = \bm \lambda_{sj1}^{l-1} + \rho \left( \bm x_{k,s}^{l} - \bm z_{s,j}^{l} \right), 
\end{equation}
\begin{equation} \label{eq21}
\bm \lambda_{sj2}^{l} = \bm \lambda_{sj2}^{l-1} + \rho \left( \bm z_{s,j}^{l} - \bm x_{k,j}^{l} \right).
\end{equation}	

Substituting \eqref{eq19} into \eqref{eq20} and \eqref{eq21}, we get 
\begin{subequations} \label{eq22}
	\begin{equation} \label{eq22a}
	\bm \lambda_{sj1}^{l} = \frac{1}{2} \left( \bm \lambda_{sj2}^{l-1} + \bm \lambda_{sj1}^{l-1} \right) + \frac{\rho}{2} \left( \bm x_{k,s}^{l} - \bm x_{k,j}^{l} \right), 
	\end{equation}
	\begin{equation} \label{eq22b}
	\bm \lambda_{sj2}^{l} = \frac{1}{2} \left( \bm \lambda_{sj2}^{l-1} + \bm \lambda_{sj1}^{l-1} \right) + \frac{\rho}{2} \left( \bm x_{k,s}^{l} - \bm x_{k,j}^{l} \right).	
	\end{equation}
\end{subequations}
By some mathematical calculation, we have $\bm \lambda_{sj1}^{l} = \bm \lambda_{sj2}^{l}$ and $\bm \lambda_{sj1}^{l} = - \bm \lambda_{js1}^{l} $, $\forall s \in \mathcal J,\, j \in \mathcal J^s$. Further, the auxiliary variable can be computed as 
\begin{equation} \label{eq23}
\bm z_{s,j}^{l} = \frac{1}{2} \left( \bm x_{k,s}^{l} + \bm x_{k,j}^{l}        \right) .
\end{equation}
Define  $\bm{\lambda}_{s}^{l-1} := \sum_{j \in \mathcal{J}^{s}} \bm \lambda_{sj1}^{l-1}$ as the local aggregate Lagrange multiplier. Then, substituting \eqref{eq23} into \eqref{eq18}, we obtain \eqref{eq24}. 

\begin{figure*}[t]
	\normalsize
	\setcounter{MYtempeqncnt}{\value{equation}}
	\setcounter{equation}{22}
	\begin{align} \label{eq24}
		\bm x_{k,s}^{l} =\left[ 2 \rho J_s \mathbf I + \mathbf H_{k,s}^{\mathsf{T}} \mathbf R_{k,s}^{-1} \mathbf H_{k,s} + \frac{1}{J} \mathbf{\Omega}_{k|k-1,s}\right]^{-1} \left[ \mathbf H_{k,s}^{\mathsf{T}} \mathbf R_{k,s}^{-1} \bm y_{k,s} + \frac{1}{J} \mathbf{\Omega}_{k|k-1,s} \hat{\bm x}_{k|k-1,s} - 2\bm{\lambda}_{s}^{l-1}  + \sum_{j \in \mathcal{J}^{s}} \rho (\bm x_{k,s}^{l-1} + \bm x_{k,j}^{l-1})\right].
	\end{align}
	\setcounter{equation}{\value{MYtempeqncnt}}
	\hrulefill
	\vspace*{4pt}
\end{figure*}
\setcounter{equation}{23}

Next, substituting \eqref{eq23} into \eqref{eq22a}, the local aggregate Lagrange multiplier can be updated as follows:
\begin{equation} \label{eq25}
\bm{\lambda}_{s}^{l} = \bm{\lambda}_{s}^{l-1} + \frac{\rho}{2} \sum_{j \in \mathcal{J}^{s}} (\bm x_{k,s}^{l-1} - \bm x_{k,j}^{l-1}).
\end{equation}
As pointed out in \cite{b14}, letting all the initial Lagrange multipliers $\bm{\lambda}_{s}^{0} $ to be $\bm 0$ for all nodes, \eqref{eq24} and \eqref{eq25} yield a convergent result after $L$ ($L$ is designed a priori) inner iterations. When $l \to  \infty$, the state estimation $\hat{\bm x}_{k|k,s} \gets \bm x_{k|k,s}^{\infty} $ on each node $s \in \mathcal{J}$ will converge to the centralized WLS estimation $\hat{\bm x}_{k|k}^{\mathrm{WLS}}$.

Next, we aim at computing the information matrix ${\mathbf \Omega}_{k|k,s}$ w.r.t  $\hat{\bm x}_{k|k,s}$ in the measurement update step. To make ${\mathbf \Omega}_{k|k,s}$ also converge to the centralized information covariance $\mathbf{\Omega}_{k|k}^{\mathrm{WLS}}$, it have the following form
\begin{equation} \label{eq26}
{\mathbf \Omega}_{k|k,s} = \left(\sum_{s=1}^{J}    \left( (\mathbf H_{k,s})^{\mathsf{T}} \mathbf R_{k,s}^{-1} \mathbf H_{k,s} \right) \right)+ {\mathbf \Omega}_{k|k-1,s}.  
\end{equation}
The summation term existing in \eqref{eq26} may cause a computational burden especially when the number of sensor nodes is large. In fact, this summation term can be computed by using the AC \cite{b4} fusion on $(\mathbf H_{k,s})^{\mathsf{T}} \mathbf R_{k,s}^{-1} \mathbf H_{k,s}$ during the inner iterations. Define the initial value at each node $s \in \mathcal J$ be $\mathbf S^{0}_{k,s} := (\mathbf H_{k,s})^{\mathsf{T}} \mathbf R_{k,s}^{-1} \mathbf H_{k,s}$. At the $t$-th iteration, each node $s$ updates its
value using the following manner
\begin{equation} \label{eq27}
\mathbf S^{t}_{k,s} = \mathbf S^{t-1}_{k,s} + \epsilon \sum_{j \in \mathcal{J}^{s}} (\mathbf S^{t-1}_{k,j} - \mathbf S^{t-1}_{k,s}).
\end{equation}
By alternatively doing this as shown in \eqref{eq27}, the values at each node $s$ will converge to the average value $\frac{1}{J} \sum_{s=1}^{J} (\mathbf H_{k,s})^{\mathsf{T}} \mathbf R_{k,s}^{-1} \mathbf H_{k,s} $.

Here, the rate parameter $\epsilon$ in \eqref{eq27} is chosen between $0$ and $\frac{1}{D_{\mathrm{max}}}$, where $D_{\mathrm{max}}$  is the maximum degree of the network $\mathcal{G}$. When the number of iterations $t \to  \infty$, \eqref{eq26} is rewritten as 
\begin{equation} \label{eq28}
{\mathbf \Omega}_{k|k,s} = J\mathbf S^{\infty}_{k,s} + {\mathbf \Omega}_{k|k-1,s}.  
\end{equation}

Since the dynamic matrices $\mathbf F_k$ and $\mathbf Q_k$ are the same for all nodes, the prediction state and information covariance is the same as its centralized counterpart in \eqref{eq7} and \eqref{eq8}, namely
\begin{equation} \label{eq29}
\hat{\bm x}_{k|k-1,s} = \mathbf F_k \hat{\bm x}_{k-1|k-1,s},
\end{equation}
\begin{equation} \label{eq30}
\mathbf{\Omega}_{k|k-1,s} = \left( \mathbf F_k \left( \mathbf{\Omega}_{k-1|k-1,s} \right)^{-1} (\mathbf F_k)^{\mathsf{T}} + \mathbf Q_k \right)^{-1} .
\end{equation}

With analysis above, the updated state $\hat{\bm x}_{k|k,s}$ and information covariance $\mathbf{\Omega}_{k-1|k,s}$ on each node $s$ are identical to the centralized fashion, when the number of iterations $l$ approaches to $\infty$. Then, at next time instant, the predicted state and information covariance are also identical to the centralized form as shown in \eqref{eq29} and \eqref{eq30}. Combining these two steps, the final estimations on each node  $s$ are same as the corresponding centralized one.

However, $l \to \infty$ is not reasonable in a practical operation. Hence, the maximum iteration $L$ is set to be a tradeoff between the performance and computational burden. The distributed WLS estimator (DWLSE) is summarized in Table \ref{table1}.
\begin{table}[t]
	\caption{Distributed WLS estimator}
	\label{table1}
	\centering
	\begin{tabular}{p{0.9\columnwidth}}
		\hline
		\textbf{Input:} For all nodes $s \in \mathcal{J}$, given $\hat{\bm x}_{0|0}$ and $\mathbf{\Omega}_{0|0}$. \\
		For each time step $k=1,...,K$, sensor node $s$ do \\
		\quad Initialization: $\mathbf S^{0}_{k,s} = (\mathbf H_{k,s})^{\mathsf{T}} \mathbf R_{k,s}^{-1} \mathbf H_{k,s}$, $\bm x_{k,s}^0=\hat{\bm x}_{k|k-1,s}$,\\ \quad $\bm{\lambda}_{s}^{0}=\bm 0$.\\
		\quad \textbf{ADMM-based consensus:}\\
		\quad For each inner iteration $l=1,...,L$ do\\
		\qquad Transmit $\bm{\lambda}^{l-1}_{sj1} $, $\bm{\lambda}^{l-1}_{sj2}$ and ${\bm x}^{l-1}_{k,s}$ to its neighboring nodes,\\  
		\qquad  Compute $\bm \lambda_{s}^{l}$ and $\bm x_{k,s}^{l}$ via \eqref{eq24} and \eqref{eq25} parallel.\\
		\quad End for\\
		\quad \textbf{Update:} Set $\hat{\bm x}_{k|k,s} \leftarrow \bm x_{k,s}^{L}$, compute $\bm \Omega_{k|k,s}$ via \eqref{eq28}.\\
		\quad \textbf{Predict:} Compute $\hat{\bm x}_{k|k-1,s}$ and $\bm \Omega_{k|k-1,s}$ via \eqref{eq29} and \eqref{eq30}.\\
		End for\\
		\textbf{Output:} Updated estimations: $\hat{\bm x}_{k|k,s}$ and $\bm \Omega_{k|k,s}$.\\
		\hline
	\end{tabular}
\end{table}

\section{Performance Evaluation}
This section compares the proposed DWLSE algorithm with DCKF \cite{b8} in a dynamic scenario. The CIF algorithm is regraded as a benchmark. The simulation is implemented in MATLAB-2019b running on a PC with processor \texttt {Intel(R) Core(TM) i7-10510U CPU 1.8GHz} and with 20GB RAM. We assess the state estimation errors by mean square error (MSE) and estimation bias between different nodes by averaged consensus estimate error (ACEE), respectively, over $M = 100$ Monte Carlo runs. 
\begin{equation*}
\mathrm{MSE} := \frac{1}{M} \sum_{m=1}^{M} \Vert \hat{\bm x}_{k|k,s}^m - {\bm x}_{k} \Vert,
\end{equation*}
\begin{equation*}
\mathrm{ACEE} := \frac{1}{J (J-1)} \sum_{s \in \mathcal J} \sum_{j \in \mathcal J} \Vert \hat{\bm x}_{k|k,s} - \hat{\bm x}_{k|k,j} \Vert,
\end{equation*}
where the superscript $m$ is the $m$-th simulation run, and $\hat{\bm x}_{k|k,s}$ is the estimation at node $s$.

 \begin{table}[t]
	\centering
	\caption{Tracker parameter settings}\label{tab:para}
	\begin{tabular}{cc}
		\hline
		Parameters & Specification \\ \hline
		Scan Time & $\mathrm{T}=1$ s \\
		Measurement Noise Cov. & $\mathbf R_{k,s} = \mathrm{diag}(200,8)$ \\
		Process Noise Cov. & $\mathbf Q_k = \mathrm{diag}(50,50,10,10)$ \\
		Initial State & $\hat{\bm x}_{1|0,s} = [20, 20,90, -80]^{\mathsf{T}}$ \\
		Initial Inf. Cov. & $\mathbf {\Omega}_{1|0,s} = \mathrm{diag} (900, 900, 16, 16)^{-1}$ \\
		State Transition Matrix & $\mathbf F_k = \begin{bmatrix}
		1  & 0 & \mathrm{T} & 0 \\ 0 & 1 & 0 & \mathrm{T} \\ 0 & 0 & 1 & 0 \\ 0 & 0 & 0 & 1 
		\end{bmatrix}$ \\
		Consensus Para. & $\epsilon = 0.65/D_{\mathrm{max}}$ \\
		Penalty Para. in DWLSE & $\rho = 0.002$ \\
		Penalty Para. in DCKF & $\mu = 100$ \\
		No. of Iterations on AC Fusion &  $10$\\
		\hline
	\end{tabular}
\end{table}
\begin{figure}[t]
	\centering
	\includegraphics[width=3.5in,height=2.2in]{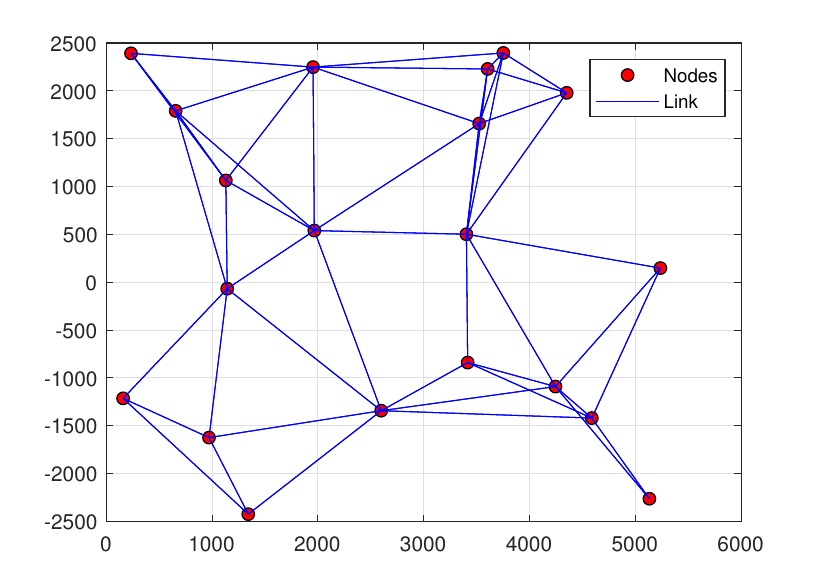}
	\caption{Sensor network with maximum communication distance $R=2$km.}
	\label{link}
\end{figure}
\begin{figure}[t!]
	\centering
	\includegraphics[width=3.5in,height=2.2in]{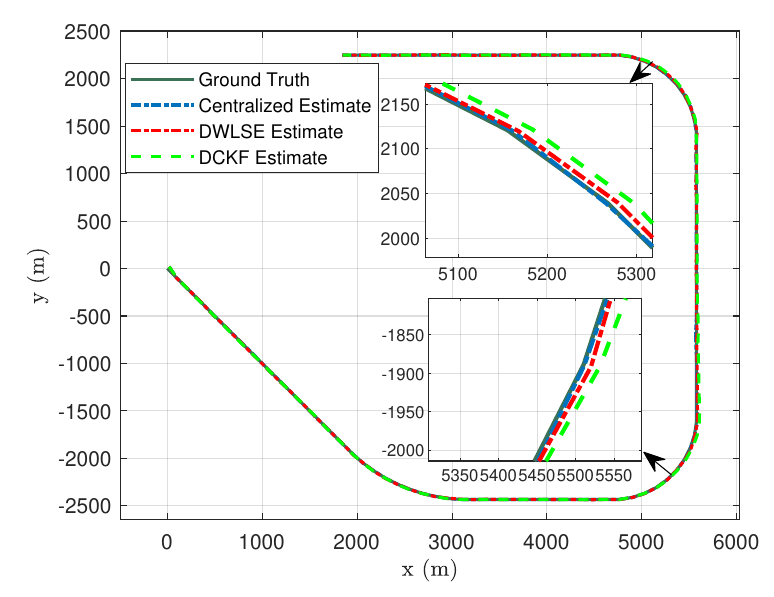}
	\caption{Ground truth and tracking results.}
	\label{result}
\end{figure}

The topology network including $20$ sensor nodes is shown in Fig. \ref{link}. The maneuvering target moves with constant velocity $\frac{5000}{36} \mathrm{m/s}$ following the trajectory as shown in Fig. \ref{result}, where the target goes through one $45^{\circ}$ and two $90^{\circ}$ turns. True initial state is $\bm x_0 = [0,0,\frac{2500\sqrt{2}}{36},\frac{2500\sqrt{2}}{36}]^{\mathsf{T}}$. Table \ref{tab:para} shows the detailed tracker parameter settings.

\begin{figure}[t]
	\centering
	\includegraphics[width=3.5in,height=2.3in]{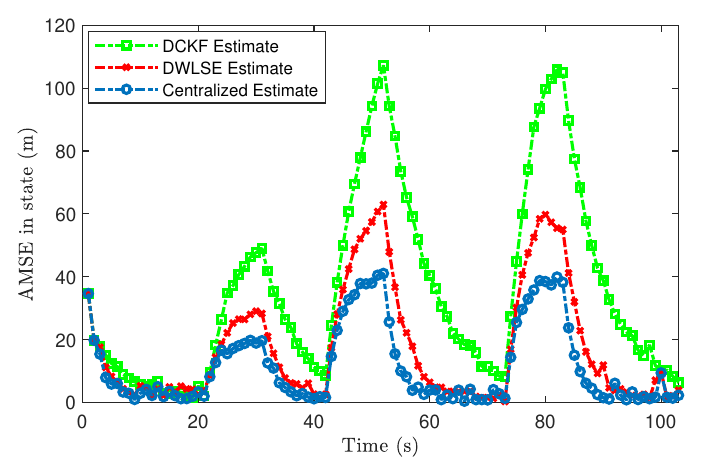}
	\caption{MSE results on node $1$ with iteration $L=20$.}
	\label{amse}
\end{figure}

\begin{figure}[t]
	\centering
	\includegraphics[width=3.5in,height=2.4in]{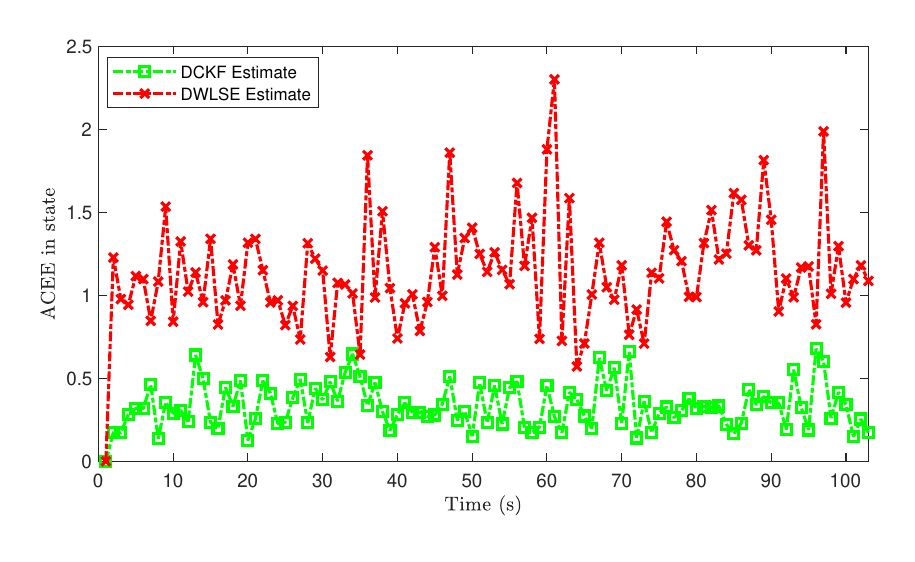}
	\caption{ACEE results between different algorithms with iteration $L=20$.}
	\label{acee}
\end{figure}

\begin{figure}[t]
	\centering
	\includegraphics[width=3.5in,height=2.3in]{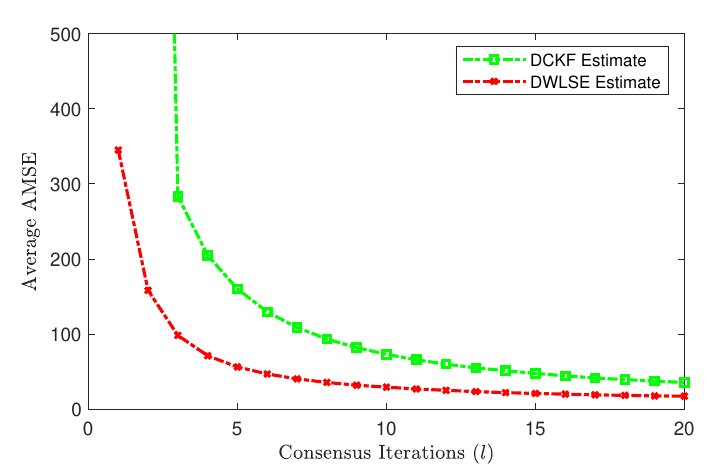}
	\caption{Average MSE with different consensus iterations.}
	\label{aveamse}
\end{figure}
\begin{figure}[t]
	\centering
	\includegraphics[width=3.5in,height=2.3in]{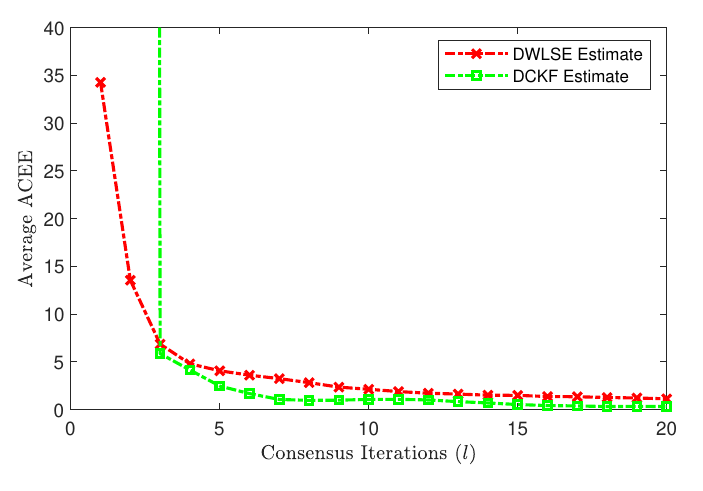}
	\caption{Average ACEE with different consensus iterations.}
	\label{aveacee}
\end{figure}

Fig. \ref{result} shows the whole tracking results for three examined filters. As expected, DWLSE has a lower estimation error than DCKF in this scenario. To obtain a more intuitive comparison results, Fig. \ref{amse} gives the MSE on node $1$ with iteration $L=20$. It is seen that DWLSE has better performance even in the three turning phases. This is because that the DWLSE constructs a novel optimization function to update the variables in parallel.

Fig. \ref{acee} shows that the differences between different nodes for DWLSE are within a reasonable range. The result demonstrates the validity of the proposed ADMM-based consensus scheme. Although the differences between nodes in DCKF are slightly small, it cannot declare that its accuracy is higher compared with DWLSE. Because when MSE metric is large, even if the ACEE value is small, it doesn't make any sense.  

The inner iteration in two examined distributed algorithms have a great impact on their performance. Thus, we check how many iterations $L$ for each distributed manner can achieve a stable behavior. Fig. \ref{aveamse} and Fig. \ref{aveacee} give the average MSE and ACEE results with different consensus iterations, respectively. 
Both DCKF and DWLSE converge to the centralized fashion when the number of iterations $l$ increases. Obviously, DWLSE attains a convergent MSE during the first $5$ iterations, while DCKF requires $10$ iterations to output an optimistic outcome. This merit makes DWLSE more appealing on a computer with limited computing capability.
Moreover, it is worth noting that DWLSE has a faster convergence rate. 
%

\section{Conclusion}
This paper proposes a distributed estimator from weighted least squares perspective. To achieve consensus between nodes, we construct an optimization function with a consensus constraint. Numerical results verity that the proposed distributed estimator gives a consensus estimations between nodes. Meanwhile, the estimations converge to the corresponding centralized form.  

\section*{Acknowledgment}
This work is supported in part by the National Natural Science Foundation of China under Grant 61873205 and Grant 61771399.


\begin{thebibliography}{99}
\bibitem{b1} R. Olfati-Saber, ``Distributed Kalman filtering for sensor networks,'' in  \textit{Proc. 2007 46th IEEE Conf. Decision and Control (CDC)}, New Orleans, LA, USA, Dec. 2007, pp. 5492-5498.
\bibitem{b2} J. Hua and C. Li, ``Distributed variational Bayesian algorithms over sensor networks,'' \textit{IEEE Trans. Signal Process.}, vol. 64, no. 3, pp. 783–798, 2016.
\bibitem{b3} G. Battistelli and L. Chisci, ``Stability of consensus extended Kalman filter for distributed state estimation,'' \textit{Automatica}, vol. 68, pp. 169-178, 2016.
\bibitem{b4} A. T. Kamal, C. Ding, B. Song, J. A. Farrell and A. K. Roy-Chowdhury, ``A generalized Kalman consensus filter for wide-area video networks,'' in  \textit{Proc. 2011 50th IEEE Conf. Decision and Control and European Control Conf. (CDC-ECC)}, Orlando, FL, Dec. 2011, pp. 7863-7869.
\bibitem{b5} M. G. S. Bruno and S. S. Dias, ``A Bayesian interpretation of distributed diffusion filtering algorithms,'' \textit{IEEE Signal Process. Mag.}, vol. 35, no. 3, pp. 118-123, 2018.
\bibitem{b6} W. Li, Y. Jia, D. Meng and J. Du, ``Distributed tracking of extended targets using random matrices,'' in  \textit{Proc. 2015 54th IEEE Conf. on Decision and Control (CDC)}, Osaka, Japan, Feb. 2015, pp. 3044-3049.
\bibitem{b7} B. Wang et al., ``Distributed fusion with multi-Bernoulli filter based on generalized covariance intersection,'' \textit{IEEE Trans. Signal Process.}, vol. 65, no. 1, pp. 242-255, 2017.
\bibitem{b8} S. Wang, H. Paul, and A. Dekorsy, ``Distributed optimal consensus-based Kalman filtering and its relation to MAP estimation,'' in \textit{Proc. 2018 IEEE International Conf. on Acoustics, Speech and Signal Process. (ICASSP)}, April 2018, pp. 3664–3668.
\bibitem{b9} A. Noroozi, M. M. Nayebi and R. Amiri, ``Iterative constrained weighted least squares solution for target localization in distributed MIMO radar,'' in \textit{Proc. 2019 27th Iranian Conf. on Electrical Engineering (ICEE)}, Yazd, Iran, Iran, May 2019, pp. 1710-1714.
\bibitem{b10} Y.~Bar Shalom, X. R. Li and T. Kirubarajan, \textit{Estimation with applications to tracking and navigation: theory algorithms and software}, John Wiley Sons press, USA, 2004.
\bibitem{b11} Q. Ling, Y. Liu, W. Shi, and Z. Tian, ``Weighted ADMM for fast decentralized network optimization,'' \textit{IEEE Trans. Signal Process.}, vol. 64, no. 22, pp. 5930–5942, 2016. 
\bibitem{b12} S. Boyd, N. Parikh, E. Chu, B. Peleato, and J. Eckstein, ``Distributed optimization and statistical learning via the alternating direction method of multipliers,'' \textit{Foundations and trends in machine learning}, vol. 3, no. 1, pp. 1–122, 2011. 
\bibitem{b13} J. Hua and C. Li, ``Distributed variational Bayesian algorithms over sensor networks,'' \textit{IEEE Trans. Signal Process.}, vol. 64, no. 3, pp. 783–798, 2016.
\bibitem{b14} J. Hua and C. Li, ``Distributed variational Bayesian algorithms for extended object tracking,'' arXiv Preprint \textit{arXiv: 1903.00182}, 2019.
\end{thebibliography}
\end{document}